\documentclass[journal]{IEEEtran}
\usepackage{cite}
\usepackage{threeparttable} 

\ifCLASSINFOpdf
   \usepackage[pdftex]{graphicx}
\else
\fi
\usepackage{amsmath}
\usepackage{color}

\begin{document}

\title{A Continuous Dual-Axis Atomic Interferometric Inertial Sensor }

\author{Pei-Qiang~Yan*, Wei-Chen~Jia*, Ke~Shen, Yue~Xin, and Yan-Ying~Feng,~\IEEEmembership{Member,~IEEE}
\thanks{This work is supported by the National Natural Science Foundation of China (Grant No. 61473166). (Corresponding author: Yanying Feng)}
\thanks{The authors are with the State Key Laboratory of Precision Measurement Technology and Instruments, Beijing, 100084, China, and A-Knows lab, Department of Precision Instruments, Tsinghua University, Beijing, 100084, China. (e-mail: yyfeng@tsinghua.edu.cn.)}
\thanks{*: Pei-Qiang Yan and Wei-Chen Jia contributed equally to this work.}}

\markboth{IEEE TRANSACTIONS ON INSTRUMENTATION AND MEASUREMENT}%
{Shell \MakeLowercase{\textit{et al.}}: A Sample Article Using IEEEtran.cls for IEEE Journals}

\IEEEpubid{}

\maketitle

\begin{abstract}
We present an interferometric inertial sensor that utilizes two counter-propagating atomic beams with transverse two-dimensional cooling. By employing three parallel and spatially aligned Raman laser beams for Doppler-sensitive Raman transitions, we successfully generate inertia-sensitive Mach-Zehnder interference fringes with an interrogation length of $2L=54\,\rm{cm}$. The sensor's capability to measure rotation and acceleration simultaneously in dynamic environments is validated through comparative analysis with classical sensors under force oscillation in different directions. Additionally, we conduct experiments on a turntable to calibrate the gyroscope's scaling factor and address nonlinearity. The angular random walk (ARW) and velocity random walk (VRW) of the senor are $3\times10^{-4}\,^\circ/\rm{\sqrt{h}}$ and $107\,\mathrm{\mu}g/\rm{\sqrt{Hz}}$, respectively, with the long-term stability reaching $9\times10^{-4}\,\rm{^\circ/h}$ for rotation and $10\,\rm{\mu g}$ for acceleration at an integration time of 1000s.
\end{abstract}


%
\IEEEpeerreviewmaketitle

\maketitle


\section{INTRODUCTION}

\IEEEPARstart{L}{ight} pulse atom interferometers(LPAI) have demonstrated remarkable measurement capabilities across various domains, including precision measurements of rotation \cite{gustavson1997,gustavson2000,durfee2006,butts2011light,tackmann2014largearea,berg2015compositelightpulse,savoie2018,avinadav2020}, acceleration \cite{mcguinness2012,lautier2014,cheiney2018}, gravity \cite{peters2001,merlet2010comparison,bidel2013,altin2013,hu2013demonstration,hauth2013first,wu2019} and gravity gradient \cite{snadden1998,sorrentino2012simultaneous,biedermann2015}. In the early stages of atomic interferometer development, the thermal atomic beam generated from an atomic oven was initially used as the matter wave source \cite{riehle1991optical,lenef1997}. With the development of laser-cooling technology, the atomic beam after transverse laser cooling achieved higher signal contrast due to a narrower transverse velocity distribution \cite{kasevich1991atomic,durfee2006}. The demonstration of pulsed cold atomic sources further reduces the velocity and velocity distribution width of atoms in three dimensions, thereby gaining advantages in interrogation time and system volume \cite{yver-leduc2003,dutta2016continuous} while enhancing signal contrast \cite{muller2007versatile,gauguet2009characterization,butts2011light,tackmann2014largearea}, leading to significant improvements in stability and sensitivity. 

In recent years, LPAI-based inertial sensors have transitioned from laboratory environments to field applications \cite{carraz2009compact,bidel2018absolute,bongs2019,lachmann2021}. For practical deployment of LPAI in field-based inertial navigation systems, a fundamental requirement is the simultaneous measurement of both rotation and acceleration by the sensor, forming the basis of an inertial measurement unit (IMU) for navigation. The atomic Mach-Zehnder interferometer, utilizing stimulated Raman transitions between hyperfine levels in a $\pi/2-\pi-\pi/2$ pulse sequence, possesses the capability to be sensitive to both rotation and acceleration. Employing two interferometers with opposite atomic velocities allows for the effective decoupling of the phase components associated with acceleration and rotation. In addition, the point-source interferometry \cite{dickerson2013multiaxisb,chen2019singlesourcea,li2021high} or multi-dimensional interferometry \cite{barrett2019multi} schemes based on cold atomic ensembles are dedicated to achieving multi-axis inertial measurements. However, interferometric inertial sensors using pulsed atomic sources face restricted measurement cycles due to the preparation of cold atomic ensembles. For inertial navigation, the sensors are expected to possess high data rate and high dynamic range of measurement, which is regarded as a primary research obstacle in employing atom interferometry in inertial sensors. Therefore, various methods have been proposed and are currently under investigation to improve data rates \cite{lautier2014,rakholia2014dualaxis,savoie2018} and eliminate dead times \cite{dutta2016continuous,savoie2018}.

The continuous counter-propagating atomic beam interferometer, offering a high data rate and eliminating dead time while canceling aliasing noise caused by Dick effect \cite{joyet2012}, presents significant advantages for dual-axis inertial sensing in dynamic environments. These ``spatial domain" interferometers, based on continuous atomic beams, eliminate the necessity for sequential laser and magnetic field control, simplifying the requirements for control devices and electro-optics.  Several studies have explored inertial-sensitive interferometers based on continuous atomic beams \cite{gustavson2000,durfee2006,xue2015,meng2024closed,sato2024closed}. Mark Kasevich's group first demonstrated an atomic interferometric gyroscope based on counter-propagating, transversely cooled cesium atomic beams and performed static measurements of the Earth's rotation rate \cite{gustavson2000}. They later enhanced the long-term stability of this gyroscope using area-reversible technology, achieving the performance requirements for high-accuracy navigation \cite{durfee2006}. T. Sato et al. further advanced the field by implementing closed-loop control in an atomic interferometric gyroscope based on a transversely cooled $^{87}$Rb atomic beam. Their experiments on a turntable confirmed that closed-loop control significantly improves the gyroscope’s dynamic range \cite{sato2024closed}.

In our previous work \cite{xue2015}, we demonstrated the first atomic interferometric gyroscope based on a continuous cold $^{87}$Rb atomic beam. Both Raman-Ramsey and Raman-Mach-Zehnder interference were observed, and the sensitivity to angular velocity was estimated. We later extended this system into an inertial sensor capable of simultaneously measuring both rotation and acceleration \cite{meng2024closed}. By implementing closed-loop control of the acceleration-induced phase shift and suppressing vibration noise, we significantly improved the long-term stability of the atomic gyroscope. Additionally, we demonstrated dynamic angular velocity measurements by benchmarking our system against a fiber-optic gyroscope. However, the system’s performance was constrained by the low atomic beam flux and the inherent complexity associated with cold atomic beams.

In this work, we present an atomic interferometric inertial sensor based on a counter-propagating, transversely cooled $^{87}$Rb atomic beam, designed for simultaneous measurement of rotation rate and acceleration. This system achieves high data rates, broad bandwidth, high sensitivity, and zero dead time. For the first time, we demonstrate dynamic dual-axis inertial measurements by directly comparing the atomic interferometric sensor with a fiber-optic gyroscope and a quartz accelerometer. To enable Doppler-sensitive Raman transitions, we employ continuous Raman laser beams separated by $L=27\ \rm{cm}$, forming a dual Mach-Zehnder interferometer configuration. Using lock-in amplifiers for orthogonal demodulation of the phase components from two interference signals, we effectively decouple the contributions of rotation and acceleration by computing their average and half-difference. Validation experiments on an air-float optical table confirm that the sensor’s measurements of rotation and acceleration align closely with those obtained from classical sensors. Additionally, we calibrate the atomic gyroscope’s scale factor using a turntable. An interferometric inertial sensor with a long-term stability of $9\times10^{-4}\,\rm{^\circ/h}$ for rotation and $10\,\rm{\mu g}$ for acceleration at an integration time of 1000s has been demonstrated. The sensor’s high precision and capability for continuous measurements with a high data update rate make it a promising candidate as a core component of a fully quantum inertial navigation unit operable in dynamic environments.

\section{METHODOLOGY AND APPARATUS}

\begin{figure*}
    \centering
    \includegraphics[width=0.98\linewidth]{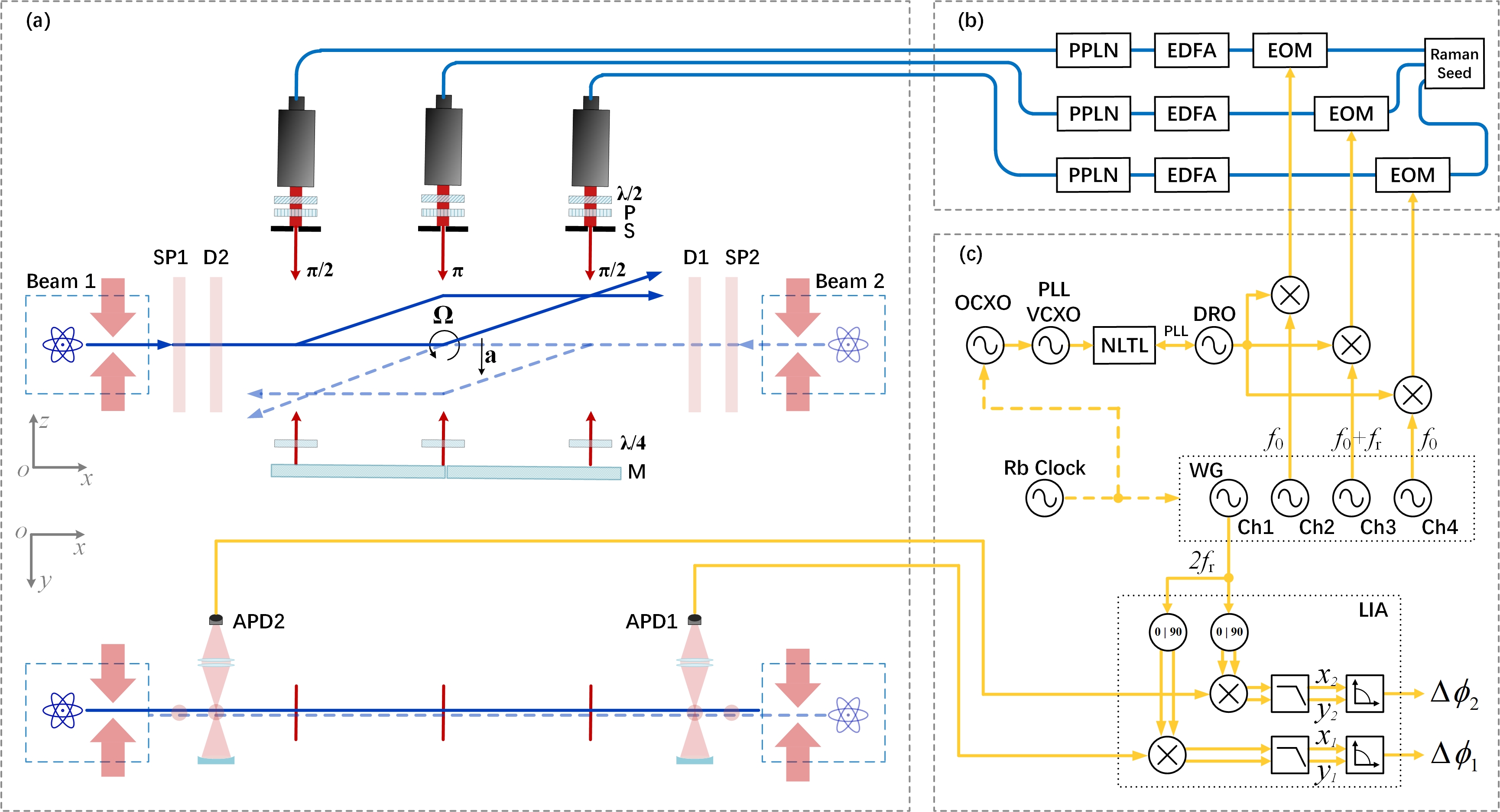}
    \caption{(a) The dual-axis atomic inertial sensor is based on two continuous counter-propagating atomic beam sources, which are generated by heating the atomic furnace and performing lateral collimation. Three spatially separated Raman light pulses with $\pi/2-\pi-\pi/2$ sequence coherently split, reflect and recombine atomic wave-packets for two atomic beams after state preparation (SP1 or SP2), and form two spatial domain atom interferometers of the Mach-Zehnder type. The interference phase shifts are measured by two APDs by collecting fluorescence induce by the detection light (D1 or D2). Two interference signals from two counter-propagating atomic beams are used for differential detection and eliminating the influence of common mode noise. The upper right corner (b) shows the Raman beams guided to the collimators comes from the modulation, amplification and SHG of the Raman seed laser.  The bottom right corner (c) illustrates the generation of RF signals and the demodulation process of interference signals. The three channels of the two waveform generators produce a frequency of $34\,\rm{MHz}$ signal mixed with $6.8\,\rm{GHz}$, which is used to modulate three Raman lasers separately, while a reference signal generated by one channel is used to demodulate the phase of the two interference signals.}
    \label{fig:configuration}
\end{figure*}

The configuration of the atomic inertial sensor is shown in Fig.~\ref{fig:configuration}. The Raman Mach-Zehnder (Raman-MZ) atomic interferometer is utilized for the simultaneous measurement of angular velocity and acceleration. Two-photon stimulated Raman transitions\cite{kasevich1991atomic} coherently manipulate atomic wave packets, with the $\pi/2$ and $\pi$ Raman lasers serving as matter-wave beam splitters/recombiners and reflectors, respectively. The atoms sequentially interact with a $\pi/2-\pi-\pi/2$ Raman laser pulse sequence, undergoing splitting, reflection, and recombination, ultimately leading to interference. The phases of the three Raman lasers are encoded by the atomic wave packets, forming the interference phase shift $\phi_{\text{MZ}}$, which contains both angular velocity and acceleration information: 

\begin{equation}
    \begin{aligned}
\phi_{\text{MZ}}&=-2\boldsymbol{k}_{\text{eff}}\cdot\left(\boldsymbol{\Omega}\times\boldsymbol{v}\right)T^2+\boldsymbol{k}_{\text{eff}}\cdot\boldsymbol{a}T^2+\phi^{(0)}.\\
    \end{aligned}
    \label{eq:MZ}
\end{equation}
The terms $\phi_\Omega=-2\boldsymbol{k}_{\text{eff}}\cdot\left(\boldsymbol{\Omega}\times\boldsymbol{v}\right)T^2$ and $\phi_a=\boldsymbol{k}_{\text{eff}}\cdot\boldsymbol{a}T^2$ represents the inertial phase shifts of rotation and acceleration, respectively. $\boldsymbol{k}_{\text{eff}}$ is the effective vector of the Raman laser, and $T=L/v$ denotes the free evolution time of the atoms between adjacent Raman lasers. The last term $\phi^{(0)}$ represents the initial phase of the interferometer, determined by the initial phase of three Raman lasers.

Aiming to decouple rotation and acceleration from Eq.(\ref{eq:MZ}), a pair of counter-propagating atomic beams are used to form dual Raman-MZ interferometers, with interference phases $\phi_{\text{MZ,1}}$ and $\phi_{\text{MZ,2}}$:
\begin{equation}
    \begin{aligned}
\phi_{\text{MZ,1}}&=-2\boldsymbol{k}_{\text{eff}}\cdot\left(\boldsymbol{\Omega}\times\boldsymbol{v}\right)T^2+\boldsymbol{k}_{\text{eff}}\cdot\boldsymbol{a}T^2+\phi_1^{(0)}.\\
\phi_{\text{MZ,2}}&=-2\boldsymbol{k}_{\text{eff}}\cdot\left(\boldsymbol{\Omega}\times\boldsymbol{v}\right)T^2-\boldsymbol{k}_{\text{eff}}\cdot\boldsymbol{a}T^2+\phi_2^{(0)}.\\
    \end{aligned}
    \label{eq:dual-MZ}
\end{equation}
Since the rotational phase shift remains the same in both $\phi_{\text{MZ,1}}$ and $\phi_{\text{MZ,2}}$, while the acceleration terms are opposite, they can be decoupled by averaging and differencing the two phase shifts $\phi_{\text{MZ,1}}$ and $\phi_{\text{MZ,2}}$:
\begin{equation}
    \begin{aligned}
\frac{\phi_{\text{MZ,1}}+\phi_{\text{MZ,2}}}{2}&=-2\boldsymbol{k}_{\text{eff}}\cdot\left(\boldsymbol{\Omega}\times\boldsymbol{v}\right)T^2+\frac{\phi_1^{(0)}+\phi_2^{(0)}}{2},\\
\frac{\phi_{\text{MZ,1}}-\phi_{\text{MZ,2}}}{2}&=\boldsymbol{k}_{\text{eff}}\cdot\boldsymbol{a}T^2+\frac{\phi_1^{(0)}-\phi_2^{(0)}}{2}.\\
    \end{aligned}
    \label{eq:decouple-MZ}
\end{equation}
Based on Eq.(\ref{eq:decouple-MZ}), the angular velocity $\Omega$ and the acceleration $a$ can be calculated.

A detailed description of the measurement method and the apparatus of the atomic sensor are provided below. For clarity, the coordinate axes in Fig.\ref{fig:configuration} are defined as follows: the x-axis aligns with the direction of the atomic beam propagation, the y-axis aligns with the direction of gravity, and the z-axis is perpendicular to both. Due to the symmetry of the configuration, we will focus on one of the atomic paths, as the other path remains identical. The atomic vapor is generated by heating the Rubidium oven to 155 $^\circ$C and is then collimated into an atomic beam as it passes through a multi-channel array. The $^{87}\rm{Rb}$ beam undergoes additional lateral collimation via a two-dimensional optical molasses. Subsequently, it enters the main chamber by passing through a porous copper pipe with a $7\,\rm{mm}$ inner diameter, effectively adsorbing divergent atoms. 

The Time of Flight (TOF) method is employed to measure the longitude velocity and the flux of atomic beams\cite{wang2024continuous,kwolek2020three}. The state-preparation laser serves as a switch to plug the atoms, and the longitude velocity distribution of the atoms can be derived from the detected fluorescence decay signal. This measurement reveals that the most probable longitudinal velocity of the atomic beam is approximately $175\,\rm{m/s}$. By integrating this velocity distribution, the flux of the atomic beam is determined to be $2.2(1)\times10^{10}\,\rm{atoms/s}$. Furthermore, by utilizing a CMOS camera for fluorescence imaging in close ($z_1=87.5\,\rm{mm}$) and distant ($z_2=700\,\rm{mm}$) windows, the 1/e diameters of the beam are measured to be $5.00\,\rm{mm}$ and $5.68\,\rm{mm}$, respectively. Combining these results, the transverse temperature of the atomic beam is approximately $524\,\mu\rm{K}$.

After the atomic beam enters the main chamber, it undergoes the processes of state preparation, Raman manipulation, and detection. Propagating along the z-axis and then reflected, the linearly polarized state preparation beam comprises two frequencies resonant with the $\left|F=2\right>\to\left|F'=1\right>$ and $\left|F=1\right>\to\left|F'=0\right>$ D2 transition lines respectively, preparing the atoms in the $\left|F=1,m_F=0\right>$ ground state. Counter-propagating Raman beams, red-detuned by $970\,\rm{MHz}$ from the $\left|F=1\right>\to\left|F'=0\right>$ transition and coupling the $\left|F=1,m_F=0\right>$ and $\left|F=2,m_F=0\right>$ states, are used to separate, deflect, and recombine these atoms. Following this, the atoms are excited by the circularly polarized detection beam, which propagates along the z-axis and is retro-reflected. The detection beam is tuned to be resonant with $\left|F=2\right>\to\left|F'=3\right>$ transition. These laser beams are generated by an integrated all-fiber optical module, then directed into collimators and expanded to the appropriate size. The induced fluorescence is collected by a pair of plano-convex lenses installed in the recessed window and a concave spherical mirror in the opposite window, achieving a theoretical collection efficiency of $2.4\,\%$ onto the avalanche photodiode (Hamamatsu C12703-01). A $\mu$-metal cylindrical shell is installed inside the main chamber as a magnetic shield, with the powered quadrupole rods providing a bias magnetic field of $B=0.67\,\rm{Gs}$ along the z-axis direction.

A $1560\,\rm{nm}$ Raman seed laser source is split into three paths using a fiber splitter (Lightcomm HPPMC-1×4-1560). Each of these paths undergoes modulation at approximately $6.834\,\rm{GHz}$ using three electro-optic modulators (iXblue MPZ-LN-10). These modulated beams are directed into an integrated device (Precilasers EFA-SSHG-780-3-CW) where power amplification via an Erbium-doped fiber amplifier (EDFA) and frequency doubling through a periodically-poled Lithium Niobate (PPLN) crystal results in a wavelength of $780\,\rm{nm}$. Due to the properties of second harmonic generation (SHG), the $6.834\,\rm{GHz}$ modulation applied at a wavelength of $1560\,\rm{nm}$ primarily generates 0th and 1st-order laser components at a wavelength of $780\,\rm{nm}$ after passing through PPLN, with a very small proportion of 2nd-order components. The Raman beams are then input into collimators with a $60\,\rm{mm}$ focal length through polarization-maintaining optical fibers.

To obtain the $6.834\,\rm{GHz}$ signal necessary for driving the electro-optic modulator, we begin with a $6.8\,\rm{GHz}$ dielectric resonator oscillator (RADITEK RDRO-A-5.0-7.9) phase-locked to a 100 MHz oscillator (Synchronization Technology STDAPM1-10-100) using a non-linear transmission line (NLTL). This $6.8\,\rm{GHz}$ signal is then split and mixed with three $34\,\rm{MHz}$ signals produced by two waveform generators (Keysight 33612A), all using single-sideband mixers (Polyphase SSB4080A). The $100\,\rm{MHz}$ phase-locked loop is referenced to a $10\,\rm{MHz}$ oven-controlled crystal oscillator (Rakon HSO14). Both the phase-locked loop and the waveform generators are referenced to a rubidium clock (SRS FS725). Afterward, each of the $6.834\,\rm{GHz}$ signals is independently power-amplified using a low-phase-noise amplifier (Connphy CLPA-6G12G-1145-S). This configuration allows for independent frequency or phase modulation of each Raman beam, providing a more flexible operational mode. 

The output signals $S_1$ and $S_2$ of the two-path atom interferometer, acquired by the APDs, can be expressed as follows:
\begin{equation}
    \begin{aligned}
    S_1&=A_1+\frac{C_1}{2}\cos\left(\phi_a+\phi_{\Omega}+\phi_0-\phi_r\right),\\
    S_2&=A_2+\frac{C_2}{2}\cos\left(-\phi_a+\phi_{\Omega}+\phi_0+\phi_r\right),
    \end{aligned}
    \label{eq1}
\end{equation}
where $A_i$ and $C_i$ represent the bias and amplitude of the interference fringes, $\phi_a$ and $\phi_{\Omega}$ are the interference phase shifts caused by acceleration and angular velocity, respectively. The phase $\phi_{0}=\phi_{0,1}-2\phi_{0,2}+\phi_{0,3}$ is introduced by the initial phase of the Raman light, while $\phi_r$ is the initial phase shift from the reflection loop influenced by mirrors’ positions.

Introducing a frequency shift $f_r$ to the EOM drive signal of the second Raman pulse ($\pi$ pulse), changes the Raman light's initial phase to:
\begin{equation}
    \begin{aligned}
    \phi'_0=&\phi_{0,1}-2[2\pi f_r(t+T)+\phi_{0,2}]+\phi_{0,3}\\
    =&-4\pi f_r t-4\pi f_r T+\phi_0,
    \end{aligned}
    \label{eq2}
\end{equation}
where $t$ is the initial moment of the Raman interaction, and $T$ is the free evolution time of the atoms between adjacent Raman pulses. Substituting Eq.(\ref{eq2}) into Eq.(\ref{eq1}) gives: 
\begin{equation}
    \begin{aligned}
    S_1&=A_1-\frac{C_1}{2}\cos(\phi_a+\phi_{\Omega}-4\pi f_r t-4\pi f_r T+\phi_0-\phi_r),\\
    S_2&=A_2-\frac{C_2}{2}\cos(-\phi_a+\phi_{\Omega}-4\pi f_r t-4\pi f_r T+\phi_0+\phi_r).
    \end{aligned}
    \label{eq3}
\end{equation}

Using the reference signal $S_{\text{ref}}=\cos(2\pi f_r t)$ for orthogonal demodulation\cite{cosens1934balance,michels1941pentode} on Eq.(\ref{eq3}), the interference phases are obtained as:

\begin{equation}
    \begin{aligned}
    \phi_{\text{MZ,1}}&=-\phi_a-\phi_{\Omega}+4\pi f_r T-\phi_0+\phi_r,\\
    \phi_{\text{MZ,2}}&=\phi_a-\phi_{\Omega}+4\pi f_r T-\phi_0-\phi_r.
    \end{aligned}
    \label{eq4}
\end{equation}

Here, the sample rate of the modulated interference signal is adopted as $469\,\rm{kHz}$. It is significantly higher than the bandwidth of the interferometer, which is evaluated as $\frac{v}{2L}=324\,\rm{Hz}$. Thus, the aliasing noise contributed to the system is negligible. A low-pass filter with cutoff frequency of approximately $20\,\rm{Hz}$ is used during orthogonal demodulation.

Combined with Eq.(\ref{eq:MZ}), the inertial phase shift can be derived as:
\begin{equation}
    \begin{aligned}
    -2\boldsymbol{k}_{\text{eff}}\cdot\left(\boldsymbol{\Omega}\times\boldsymbol{v}\right)T^2&=-\frac{\phi_{\text{MZ,1}}+\phi_{\text{MZ,2}}}{2}+4\pi f_r T-\phi_0,\\
    \boldsymbol{k}_{\text{eff}}\cdot\boldsymbol{a}T^2&=-\frac{\phi_{\text{MZ,1}}-\phi_{\text{MZ,2}}}{2}+\phi_r.
    \end{aligned}
    \label{eq5}
\end{equation} 

Thus, the rotational rate $\Omega$ and acceleration $a$ can be computed. Due to the presence of $\phi_0$, $\phi_r$ and $f_r$, the output inertial signals may exhibit biases, thus requiring further calibration and correction work.

In spatial-domain interferometers, it is common practice to shape the Raman beam into a narrow ellipse to enhance performance \cite{gustavson2000,kwolek2022continuous}. Along the direction of the atomic beam's longitudinal velocity, the waist of the Raman light determines the range of transverse velocities over which the atoms can interact. This range is constrained by the Doppler shift and plays a crucial role in influencing the interferometer's signal contrast. However, if the waist is too narrow, it can reduce the Rayleigh range below the transverse size of the atomic beam, leading to significant wavefront curvature and spatially uneven distribution of the Gouy phase shift, thereby inducing non-uniform Raman phase on the atoms. Additionally, it's essential for the Raman beam to have sufficient width in the transverse direction of the atomic beam. This ensures a uniform intensity and a flat wavefront, contributing to the interferometer's overall performance.

To achieve a large interference fringe amplitude, the optimal Raman beam waist ($2\omega_0$) along the x-axis is approximately $100\sim200\,\mu\rm{m}$, corresponding to a Rayleigh range ($z_R$) of $10\sim40\,\rm{mm}$. Simultaneously, the waist of the Raman beam along the y-axis should be larger than the measured transverse atom beam size (rms) $\sigma_{\rm{atom-y}}=8\,\rm{mm}$. Obtaining such a Raman beam through optical shaping from an optical fiber port using conventional approaches, such as cylindrical lenses, requires precise adjustment, alignment, and the correction of beam pointing deviations introduced by the wedge in the non-focusing axis of the cylindrical lenses. In addition to the beam shape requirements, it is crucial to ensure the mutual parallelism of the three Raman beams and their reflection beams with an accuracy better than $35\,\mu\rm{rad}$. This can be calculated using the formula $\Delta\theta_R<\pi/(\sigma_{\rm{atom-z}}k_{\rm{eff}})$, where $k_{\rm{eff}}$ is the two-photon wave vector of the Raman beams, and $\sigma_{\rm{atom-z}}$ is the rms transverse size of the atomic beam along the z-axis \cite{kwolek2022continuous}. Due to the strict requirements imposed on the optical design and parallelism adjustment when dealing with narrow-ellipse beams, we decide to use $\phi\,11\,\rm{mm}$ collimated Raman beams and have proposed the following optical scheme which is easier to operate.

Elongated reflecting mirrors serves as reference benchmarks to achieve precise parallel alignment of the collimated Raman beams. Given the challenge of achieving high flatness and low curvature on a single elongated mirror exceeding $2L=54\,\rm{cm}$, we have opted to customize two microcrystalline mirrors with a surface flatness of $\lambda/10$ and surface parallelism better than $10\,\mu\rm{rad}$, allowing us to align two pairs of Raman $\pi/2-\pi$ beams separately, creating a relay-like configuration.

When the Raman beam emitted from the fiber collimator aligns perpendicularly with the mirror surface, it reflects back into the collimator, allowing some of its power to couple into the fiber. According to optical simulations, when maintaining at least $90\%$ of the optimal power coupling efficiency, the incident angle between the incoming laser beam and the mirror will not exceed $7\,\mu\rm{rad}$. This precision in angle adjustment and the quality of elongated mirrors guarantee a worst-case scenario of $24\,\mu\rm{rad}$ in parallelism between any two Raman beams, meeting the calculated requirements. For this alignment, a 50:50 fiber splitter is employed to guide the Raman beam through the main cavity, directing it towards the mirror. The adjustment of its pointing is performed while simultaneously monitoring the laser power coupled back into the splitter.

The configuration of the Raman optical path is shown in Fig.~\ref{fig:configuration}(a). The Raman $\pi$ beam is adjusted to have an angle of approximately $\theta=0.3^\circ$ with respect to the z-axis in order to sufficiently separate the Doppler-sensitive transition peaks with a frequency intervals $\Delta f=\frac{2k_{\text{eff}}v\sin(\theta)}{2\pi}=4.7\,$MHz. To calibrate the horizontal orientation of the beam and eliminate the influence of gravity projection on inertial phase shift, a pentagonal prism and a water surface (not shown in the figure) are utilized to reflect and couple the Raman beam back into the optical fiber. Two elongated mirrors are individually adjusted to couple a portion of the Raman $\pi$ beam back into the fiber, resulting in coupling efficiencies of $28\%$ and $2\%$ due to their intentional asymmetric placement. Finally, the parallel alignment of the three Raman beams is achieved by adjusting the self-coupling of the two Raman $\pi/2$ beam collimators, which have efficiencies of $54\%$ and $33\%$.

Waveplates and polarizers are installed along each individual Raman optical path, and their orientations are fine-tuned to achieve the lin$\perp$lin configuration for suppressing the Doppler-free transitions. Subsequently, a 1 mm-wide slit is added to each Raman optical path to transform the circular Gaussian beams into 1 mm-wide elongated beams. The slit's position is adjusted to center it on the Raman spot using a powermeter. Further refinement of the slit's position can be achieved by optimizing the contrast of the interference fringes.

\begin{figure}
    \centering
    \includegraphics[width=0.98\linewidth]{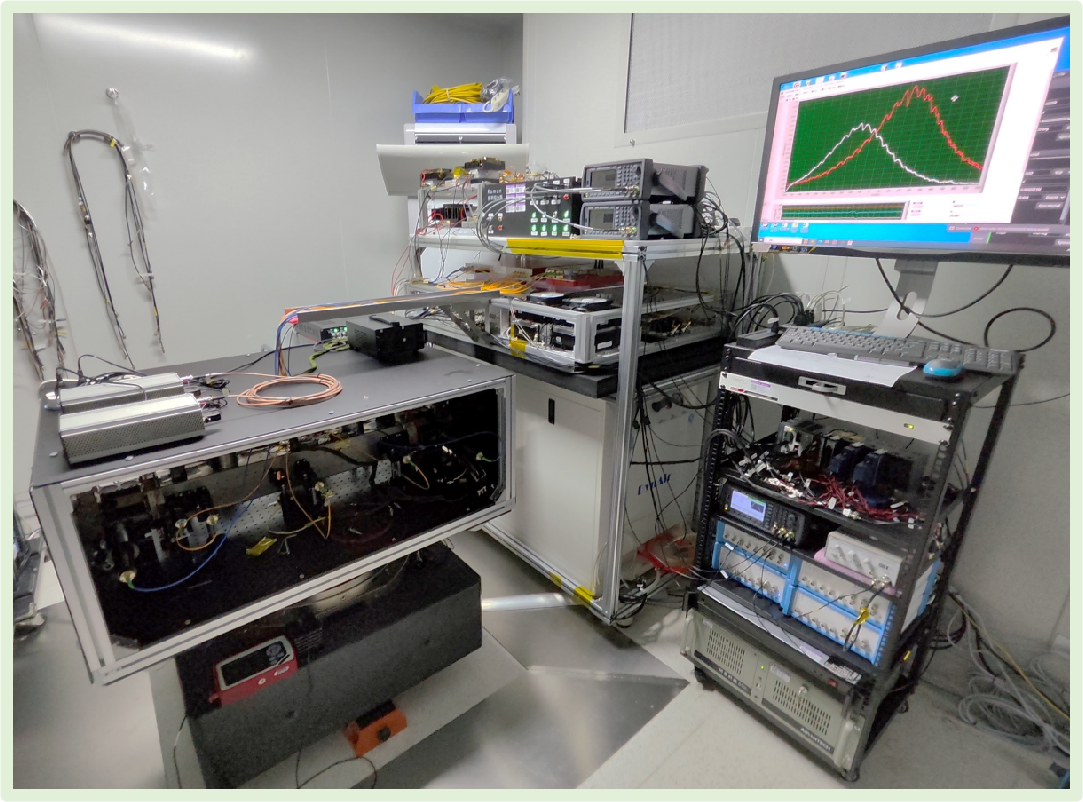}
    \caption{The photo of the atomic interferometric inertial sensor, including a sensor head, along with the corresponding laser, microwave, and control units.}
    \label{fig:photo}
\end{figure}

Finally, we have developed a prototype interferometric sensor, which contains a sensor head, and the corresponding laser, microwave and control unit, as shown in Fig. \ref{fig:photo}. This configuration yields a compact overall size of $1.1\,\rm{m}(L)\times0.8\,\rm{m}(W)\times0.6\,\rm{m}(H)$, enabling it to readily adapt to dynamic testing conditions.

\section{INTERFERENCE}

To generate the interferometric fringes, a relative frequency offset $f_r=200\,\rm{Hz}$ is introduced to the $34\,\rm{MHz}$ signal of the Raman $\pi$ beam compared to the Raman $\pi/2$ beam. This offset causes the atomic interference phase to ramp at a rate of $\dot\phi=4\pi f_R$ \cite{gustavson2000}. The typical contrasts of these two interference fringes are approximately $C_1=5.9(2)\,\%$ and $C_2=3.7(2)\,\%$, which are calculated from $C_{i}=(V_{max,i}-V_{min,i})/(V_{max,i}+V_{min,i}) (i=1,2)$, respectively. The values of $V_{max}$ and $V_{min}$ for each interference signal are obtained through the sinusoidal fitting. The contrasts are influenced by the velocity selection effect of Raman transition processes. Specific factors include the characteristics of the atomic beams, the linewidth and parallelism of the Raman lights, and the angles between the Raman lights and the atomic beams. Misalignment of the dual atomic beams in our system may generate different angles with respect to the Raman lights, primarily causing contrast differences. Additionally, differences in the transverse temperatures and mean longitudinal velocity of each atomic beam interferometer are subject to slight variations, which can result in different contrasts. Differences in detection noise between the two interferometers could also contribute to the observed contrast disparity. These signals are then fed into digital lock-in amplifiers (Zurich Instruments MFLI), where they undergo demodulation using a common reference signal at $2f_r=400\,\rm{Hz}$. To effectively filter out the modulation frequency while retaining the capability to measure dynamic signals, a $20\,\rm{Hz}$ bandwidth is selected for the low-pass filters. Thanks to the lock-in detection method, the demodulated channels can simultaneously provide the sine, cosine, phase and amplitude information.

\begin{figure}
    \centering
    \includegraphics[width=1.0\linewidth]{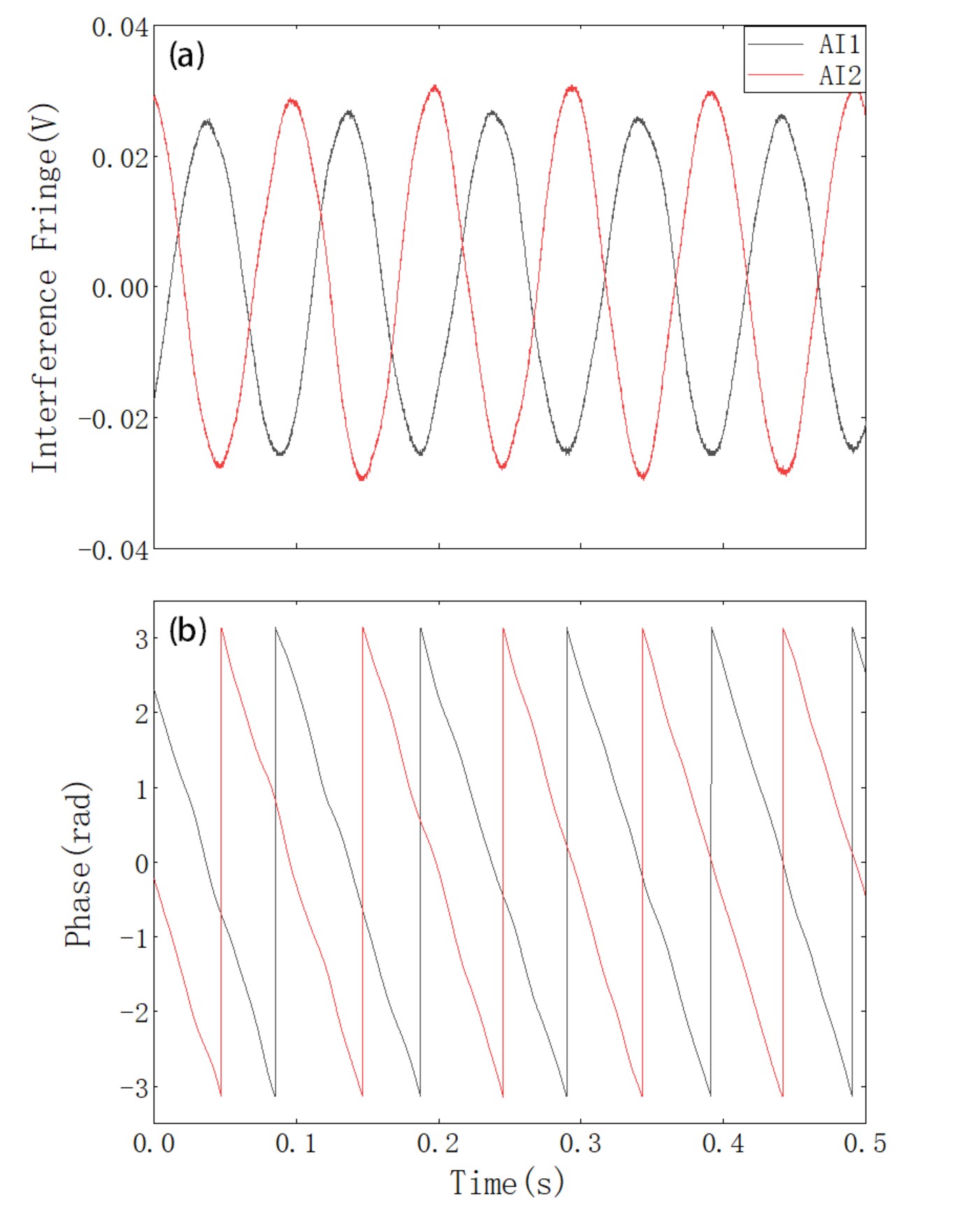}
    \caption{The demodulated interference fringes (a) and phase (b) from lock-in amplifiers when introducing phase modulation to the $34\,\rm{MHz}$ signal of the first Raman $\pi/2$ beam.}
    \label{fig:fringes}
\end{figure}

At this stage, by introducing an additional phase modulation to the $34\,\rm{MHz}$ signal of the first Raman $\pi/2$ beams, the demodulated sine component and phase of two interferometers can be observed as illustrated in Fig.~\ref{fig:fringes}. The demodulated phase distinctly reveals the linear sweep of the interference phase, resulting in a saw-tooth wave shape with phase jumps occurring due to the $\pm\pi\,\rm{rad}$ limitation.

Without phase modulation to Raman $\pi/2$ beam, we record the demodulated signal over a $100\,\rm{s}$ period with a sample rate of $100\,\rm{Hz}$, which is set to 5 times the cutoff frequency of the interference phases to further suppress potential aliasing noise. The phase noise of each interferometer is calculated to be $0.44\,\rm{rad/sample}$ and $0.45\,\rm{rad/sample}$. The corresponding rotation rate and acceleration components of the phase noise are $0.03\,\rm{rad/sample}$ and $0.44\,\rm{rad/sample}$, resulting in rotation and acceleration sensitivities of $0.25\,\mu\rm{rad/s}/\sqrt{Hz}$ and $0.12\,\mathrm{m}g/\rm{\sqrt{Hz}}$. Based on Eq.(\ref{eq5}), the acceleration measurement has an advantage in suppressing common-mode noise compared to the angular velocity measurement. However, the current setup is experiencing significant environmental vibration noise along the acceleration-sensitive axis, which introduces phase noise to $\phi_r$ by affecting the mirror mounts, leading to a worse short-term sensitivity of acceleration measurement than our expectations. 

Aiming to reveal the intrinsic noise level of the sensor under vibration-insensitive condition, the sensor's performance are evaluated in the spin-echo configuration\cite{kwolek2022continuous}. The sensitivities are measured to be $70 \ \rm{nrad/s/\sqrt{Hz}}$ and $1.5\ \rm{\mu g/\sqrt{Hz}}$. It achieves approximately a threefold improvement in rotational sensitivity and nearly two orders of enhancement in acceleration sensitivity. Thus, by incorporating a vibration compensation algorithm that uses a classical accelerometer in conjunction with the atomic inertial sensor, or by providing an experimental environment with reduced vibration noise in future, we expect to significantly improve the sensitivity of the atomic sensor. 

The sensitivity of the atomic interferometer is ultimately limited by its shot noise, which can be calculated as:
\begin{equation}
    \begin{aligned}
    \Delta\Omega_{\text{shot}}&=\frac{v}{2C\sqrt{N}k_{\text{eff}}L^2},\\
    \Delta a_{\text{shot}}&=\frac{v^2}{C\sqrt{N}k_{\text{eff}}L^2},
    \end{aligned}
    \label{eq:shot-noise}
\end{equation} 
where $C$ represents the fringe contrast, and $N$ denotes the flux of the atomic beam. Based on the current parameters, the shot-noise-limited sensitivities are calculated as $\Delta \Omega_{\text{shot}}=11\ \rm{ nrad/s/\sqrt{Hz}}$ and $\Delta a_{\text{shot}}=370\ \rm{ng/\sqrt{Hz}}$, which is approximately 4$\sim$7 times better than the current intrinsic noise level of the sensor. The limited intrinsic sensitivity of the sensor is primarily attributed to the technical noise, such as laser frequency, power and initial phase noise. A series of experimental evidences demonstrate the influence of several of these factors. For instance, by applying a fast modulation to the laser frequency, its transfer coefficients to the inertial outputs are measured to be approximately $c_\Omega = \delta \Omega/\delta f = 11\ \rm{nrad/s/kHz}$ and $c_a = \delta a/\delta f = 1.1\ \rm{\mu g/kHz}$, respectively. By combining these coefficients with the noise spectral density of the laser frequency, its contribution to the interferometer sensitivity can be further evaluated.

Moreover, cross-coupling between rotational and accelerational measurements occurs when the atomic beam velocities differ, affecting sensitivity. To assess this, we theoretically estimate the impact, assuming identical most probable velocities ($v_0$) and decoupling inertial phase shifts via averaging and differential methods while neglecting initial phase shifts $\phi_0$ and $\phi_r$. The resulting expressions are:
\begin{equation}
\begin{split}
&\phi_{\mathrm{avr}}=\frac{\phi_{\text{MZ,1}}+\phi_{\text{MZ,2}}}{2}=-\frac{2k_\mathrm{eff}\Omega^{^{\prime}}L^2}{v_0}, \\
&\phi_{\mathrm{diff}}=\frac{\phi_{\text{MZ,1}}-\phi_{\text{MZ,2}}}{2}=\frac{k_\mathrm{eff}a^{\prime}L^2}{v_0^2},
\end{split}
\label{eq:atomsv1}
\end{equation}
where $\Omega'$ and $a'$ represent the measured values of angular velocity and acceleration.

However, when the velocity difference between the two atomic beams is considered (denoted as $v_1$ and $v_2$), the resulting interference phase shift can be expressed as:
\begin{equation}
\begin{split}
\phi_{\text{MZ,1}}=-\frac{2k_\mathrm{eff}\Omega L^2}{v_1}+\frac{k_\mathrm{eff}aL^2}{v_1^2},\\
\phi_{\text{MZ,2}}=-\frac{2k_\mathrm{eff}\Omega L^2}{v_2}-\frac{k_\mathrm{eff}aL^2}{v_2^2},
\end{split}
\label{eq:atomsv2}
\end{equation}
where $\Omega$ and $a$ represent the actual angular velocity and acceleration, respectively. Assuming the velocities of the two atomic beams are given by $v_1=(1+\epsilon)v_0$ and $v_2=(1-\epsilon)v_0$, where $\epsilon$ denotes a coefficient characterizing the velocity difference between the two atomic beams, the power spectral density (PSD) of their measurement noise can be calculated as:
\begin{equation}
    \begin{split}
   S_{\Omega'}(f) & =(\frac{1}{1-\epsilon^2})^2S_\Omega(f)+(\frac{\epsilon}{\left(1-\epsilon^2\right)^2})^2\frac{S_a(f)}{v_0^2},\\
    S_{a'}(f) & =(\frac{1+\epsilon^2}{\left(1-\epsilon^2\right)^2})^2S_ a(f)+(\frac{2\epsilon}{1-\epsilon^2}v_0)^2S_\Omega(f).
    \end{split}
    \label{eq:atomsv3}
\end{equation}
Here, $S_{\Omega'}(f)$ and $S_{a'}(f)$ represent the noise PSDs of the decoupled rotation and acceleration measurements, respectively, while $S_{\Omega}(f)$ and $S_{a}(f)$ denote the actual noise PSDs without cross-coupling. The $S_{\Omega}(f)$ and $S_{a}(f)$ are assumed to be statistically independent. 

For a velocity consistency of $\epsilon = 1\%$, substituting the experimental sensitivities of $0.25\,\rm{\mu rad/s/\sqrt{Hz}}$ and $0.12\,\rm{mg/\sqrt{Hz}}$ into Eq. (\ref{eq:atomsv3}) gives actual sensitivities of $0.24\,\rm{\mu rad/s/\sqrt{Hz}}$ and $0.12\,\rm{mg/\sqrt{Hz}}$ without cross-coupling. Due to lower acceleration sensitivity, velocity mismatch causes a $4\%$ deviation in rotational sensitivity, while cross-coupling from rotation to acceleration remains negligible. A more precise estimation requires experimental validation. Cross-coupling can be further suppressed by optimizing atomic oven temperature control and transverse collimation optics to enhance velocity consistency and trajectory alignment, as well as implementing closed-loop control to stabilize the interferometer phase at zero, reducing sensitivity to atomic longitudinal velocity \cite{sato2024closed}.

\section{DYNAMICS}

\begin{figure*}
    \centering
    \includegraphics[width=0.98\linewidth]{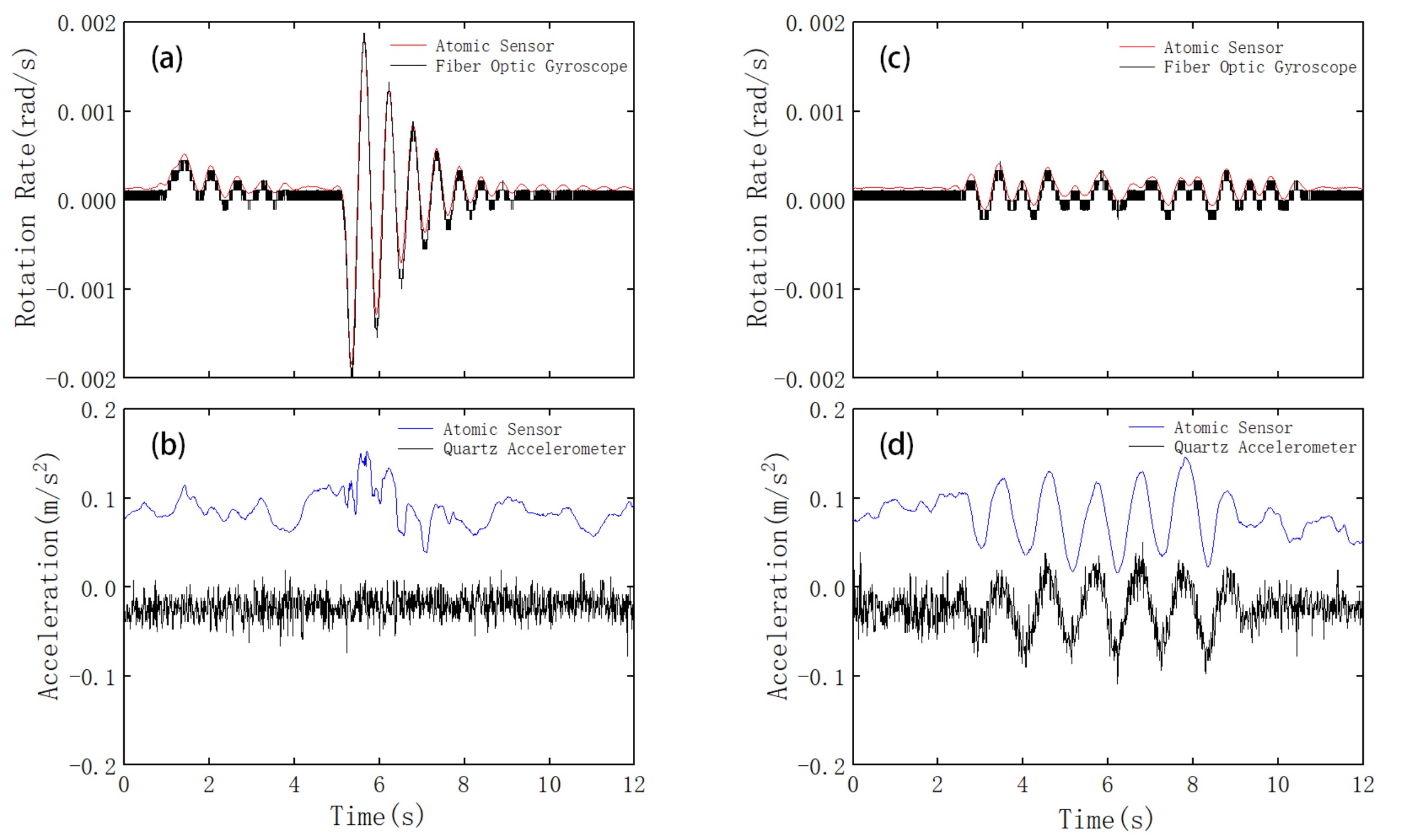}
    \caption{Comparison of the atomic sensor with the fiber optic gyroscope and the quartz accelerometer. The yaw and pitch motions of the optical platform are generated by manual pushing, resulting in signals (a)(b) and (c)(d) respectively. Remarkable agreement is observed with the fiber optic gyroscope, while the bias between the atomic sensor and quartz accelerometer may be attributed to potential misalignment in their sensitive axes.}
    \label{fig:comparison}
\end{figure*}

The ambiguity dynamic range of inertial measurement based on interference fringes is constrained by half a cycle between any adjacent peaks and valleys. In the case of a lock-in amplifier with two orthogonal channels, the quadrature demodulation technique can directly extract the interferometric phase, thereby expanding the ambiguity phase range to $2\pi$ as shown in Fig.~\ref{fig:fringes}(b). Despite the interference phase shifts resulting from dynamic motion may exceed it, compensation for these phase shifts is effectively achieved by adding or subtracting $2\pi\,\rm{rad}$ when the phase suddenly jumps from $\pm\pi$ to $\mp\pi$. This is possible due to the damping effect of the vibration isolators we used, which enables a gradual change in rotation rate or acceleration. As a result, the sensor's dynamic range can be extended to multiple cycles of $2\pi$, significantly surpassing the capabilities of the original method. After decoupling the rotational and accelerational phase shifts from the average and half-difference of two interferometric phases, we convert them into inertial quantities using theoretically derived scaling factors. 

To validate the dynamic response capability of the atomic inertial sensor, an optical platform equipped with four pneumatic vibration isolators (Newport S-2000) is utilized as the testing platform. A comparison is made between the sensor's output and that of classical inertial sensors under artificial motion excitation. A fiber optic gyroscope (BATOT FOG-1120H), featuring a digital output with a data rate of 256 Hz and a resolution of $1.1\times 10^{-4}\ \rm{rad/s}$, is mounted at the center position of the atomic sensor's baseplate, with its sensitive axis for rotational measurements aligned in the same direction as the atomic sensor, along the vertical direction. A quartz flexural accelerometer (ASIT JN-06A) is installed directly above the center of the main vacuum chamber of the atomic sensor, with its sensitive axis for acceleration measurements aligned horizontally. It is positioned $120\,\rm{mm}$ above the vertical axis of the Raman $\pi$ beam. Pushes in different directions and positions are applied to the optical platform, inducing yaw and pitch motions.

To validate the rotational measurement, a yaw motion is induced by gently pushing and holding one corner of the optical platform away from the atomic sensor. Then, we release it rapidly to allow free motion. Due to the lateral restoring forces and damping of the vibration isolators, the optical platform undergoes oscillatory rotational motion along its vertical axis, which gradually attenuates. Since the inertial sensor is positioned at one end of the optical platform rather than at the center, under conditions of small angular motion, the oscillatory motion of the atomic sensor can be approximated as a combination of rotation about its vertical axis and translation along the x-axis.

The output signals from the atomic gyroscope and the fiber optic gyroscope exhibit good consistency as shown in Fig.~\ref{fig:comparison}(a). Due to the challenging task of manually controlling the pushing force, the inertial signal shows several ripples in the positive rotational direction during the initial 4 seconds. After maintaining stationarity for about 1 second, the optical platform is rapidly released for free motion. The waveform of the oscillation process of the inertial sensor is fitted using a second-order system model, resulting in a resonant frequency of $2.13(3)\,\rm{Hz}$ and a damping ratio of 0.59(1). The acceleration output of the atomic sensor in Fig.~\ref{fig:comparison}(b) exhibits some low-frequency fluctuations and experiences more pronounced changes between the 5th and 6th seconds due to the release of the platform.

When validating the acceleration measurement, it is challenging to smoothly push the optical platform along the direction of the acceleration-sensitive axis. Instead, we modulate the acceleration by changing the angle between the acceleration-sensitive axis and the direction of gravity. This is achieved by applying periodic downward pressure and release to one end of the optical platform, causing the platform to pitch about its horizontal short axis. As mentioned earlier, the atomic sensor also experiences translation along the gravity axis. As shown in Fig.~\ref{fig:comparison}(d), the acceleration waveforms measured by the atomic inertial sensor and the quartz accelerometer are relatively consistent during the time interval from 2 to 10 seconds.

When comparing the atomic sensor with two classical sensors, the measured rotation and acceleration biases under near-static conditions are approximately $130\ \rm{\mu rad/s}$ and $0.08\ \rm{m/s^2}$, respectively, which deviate significantly from the expected Earth's rotation projection at the testing location ($40^{\circ}\rm{N}$, $\Omega_\mathrm{E}\sin(40^{\circ})\approx47\ \rm{\mu rad/s}$) and the expected horizontal static acceleration ($0\ \rm{m/s^2}$). These discrepancies arise from several factors. First, the biases can partly be explained by Eq.(\ref{eq5}) derived in Section II. The interferometer phase modulation introduces a rotational phase shift bias of $4\pi f_r T$ ($\sim3.88\ \rm{rad}$), corresponding to an angular velocity bias of about $289\ \rm{\mu rad/s}$. Additionally, uncertainties in the initial Raman phase $\phi_0$ and Raman reflection phase $\phi_r$ contribute to the observed biases. The uncertainty in $\phi_0$ originates from the inability of two signal generators to achieve precise phase synchronization leading to a random phase offset within $(-\pi,\,\pi)$ radians each time the system is powered on, resulting in an indeterminate angular velocity bias within $(-234,\ 234)\ \rm{\mu rad/s}$. Similarly, the uncertainty in $\phi_r$, influenced by the positioning of the three reflective surfaces, falls within $(-\pi,\ \pi)$ radians due to measurement limitations, leading to an acceleration bias of $(-0.16,\ 0.16)\ \rm{m/s^2}$. Second, despite the application of a phase unwrapping method to correct phase jumps, ambiguity persists in determining the initial cycles of $\phi_{\rm{MZ,1}}$ and $\phi_{\rm{MZ,2}}$, introducing uncertainties in $\phi_{\Omega}$ and $\phi_a$ that manifest as biases of $(k \times 234)\ \rm{\mu rad/s}$ for angular velocity and $(k \times 0.16)\  \rm{m/s^2}$ for acceleration, where $k = 0, \pm1, \pm2, \dots$. Third, the tilt angle of the Raman beams is minimized to less than $5\ \rm{\mu rad}$, resulting in a negligible gravitational projection bias of less than $5 \times 10^{-5}\ \rm{m/s^2}$ and an angular velocity bias of $\left|\Omega_{\mathrm{E}} \sin(40^{\circ}) - \Omega_{\mathrm{E}} \sin(40^{\circ}+\theta) \right| < 1\ \rm{nrad/s}$, making alignment-related biases insignificant in the current system. Finally, system errors arising from inaccurate physical parameters and long-term drift of the atomic sensor further contribute to these deviations. In summary, uncertainties in the Raman phase and interference phase cycle ambiguity significantly impact the inertial bias errors, necessitating further experimental investigations.For clarity, these bias estimations and potential correction strategies are summarized in Table I.

\begin{table*}[htbp!]
    \centering
    \begin{threeparttable}
    \caption{The bias sources and correction strategies of the sensor.}
    \begin{tabular}{c| c| c |c}
    \hline
         Bias sources & Bias($\Omega,\rm{\mu rad/s}$) & Bias($a,\rm{m/s^{2}}$) & Correction Strategy\\ \hline
         Phase Modulation &$\sim289$&--- & Direct compensation \\ \hline
         Initial phase $\phi_0$ & $(-234,234)$ & --- & Measurement and active stabilization\\
        \hline
        Initial phase $\phi_r$ & --- & $(-0.16,0.16)$ & Measurement and active stabilization\\
        \hline
         Ambiguity in Phase Cycles& $k\times 234$ & $k\times0.16$ &Phase unwrapping and closed-loop measurement \\
        \hline
         System error and drift & \multicolumn{2}{|c|}{Under Evaluation} & Long-term stability improvement\\
        \hline
         Alignment-related bias& $<0.001$ & $<5\times10^{-5}$ &Temporarily negligible\\
         \hline
         Experimental results & 130 & 0.08 & \\
         \hline
         Theoretical values $^*$ & 47 & 0  & \\
         \hline
    \end{tabular}
    \begin{tablenotes}
        \footnotesize
        \item$^*$ Theoretical value is calculated under ideal alignment conditions, where the projection of gravity along the effective Raman wave vector is zero.
    \end{tablenotes}
    \label{tab:comparation}
    \end{threeparttable}
\end{table*}

The atomic inertial sensor is mounted on a turntable (D600, ABTECH-space) for further testing, which has an angular velocity resolution of less than $0.0004\ \rm{^\circ/s}$, an end runout of less than 1.5$^{\prime\prime}$, and a maximum rotational speed of up to $60\ \rm{^\circ/s}$. The turntable is manually adjusted to gradually increase the rotation speed within the range of $\pm\,0.03 \ \rm{^\circ\rm{/s}}$. Both clockwise and counterclockwise rotations are conducted at each speed setting, each lasting for 30 seconds, with brief pauses between each rotation stage. The rotation phase shift between the 11th and 25th seconds in each 30-second rotation stage is selected for averaging phase calculations.

\begin{figure}
    \centering
    \includegraphics[width=0.95\linewidth]{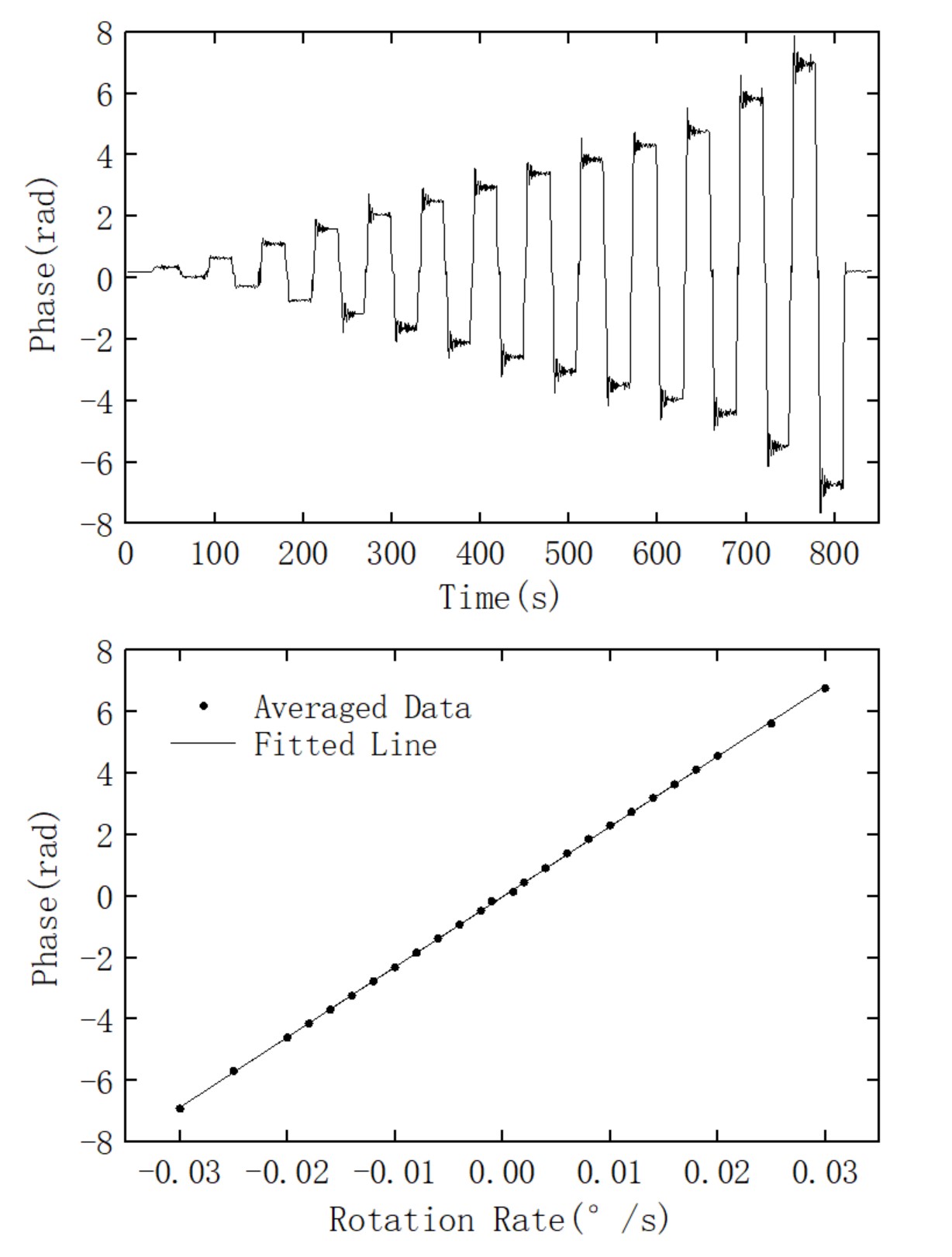}
    \caption{(a) The rotational phase shift of the atomic sensor occurs as the rotation rate of the turntable continuously increases in alternating clockwise and counterclockwise directions. (b) Linear fitting for the average phase shifts at various rotation rates.}
    \label{fig:scalefactor}
\end{figure}

The scatter plot of turntable rotation speed and the rotational phase output of the atomic sensor is fitted, resulting in the atomic gyroscope’s scaling factor $K_m=228.4\,\rm{rad/(^\circ/s)}$ and an output nonlinearity of $1.2\,\%$ across the range of $\pm\,0.03^\circ\rm{/s}$. The scale factor shows a relative error of $2.5\,\%$ compared to the theoretical value of $K_t=2k_{\rm{eff}}L^2/v=234.2\,\rm{rad/(^\circ/s)}$, primarily due to variations of atomic velocities during the measurement process. The fluctuation in atomic velocity also contributes to the calibration nonlinearity by affecting the scale factor. Additionally, nonlinearity may arise from the turntable's angular velocity resolution, especially at lower angular velocity measurement points, as well as from its instability during motion. The output of the atomic gyroscope exhibits damped oscillations in each rotation stage, and these oscillations become more pronounced with increasing rotation speed. This may be attributed to various factors, including the system placed on the air-bearing turntable not being properly balanced and the drive controller parameters of the turntable’s drive motor not being optimized for this weight. Despite the observed oscillations, the present dynamic range of $\pm\,0.03^\circ\rm{/s}$ has substantially exceeded the $\pm\,0.013^\circ\rm{/s}$ range (equivalent to an ambiguity phase of $\pm\,\pi\,\rm{rad}$) thanks to the phase jump compensation. However, it is still constrained by vibration noise and a reduction in signal contrast during turntable rotation, which can be addressed through closed-loop measurement in future studies. In closed-loop operation, the interference phase remains locked to the setpoint through real-time feedback control of the Raman laser parameters, and the inertial measurement results are calculated from the control signal. By applying appropriate control methods, the scale factor of the closed-loop sensor can become independent of atomic velocity\cite{sato2024closed}, which suppresses the instability induced by the velocity variation and helps mitigate the nonlinearity.

\begin{figure*}
    \centering
    \includegraphics[width=0.98\linewidth]{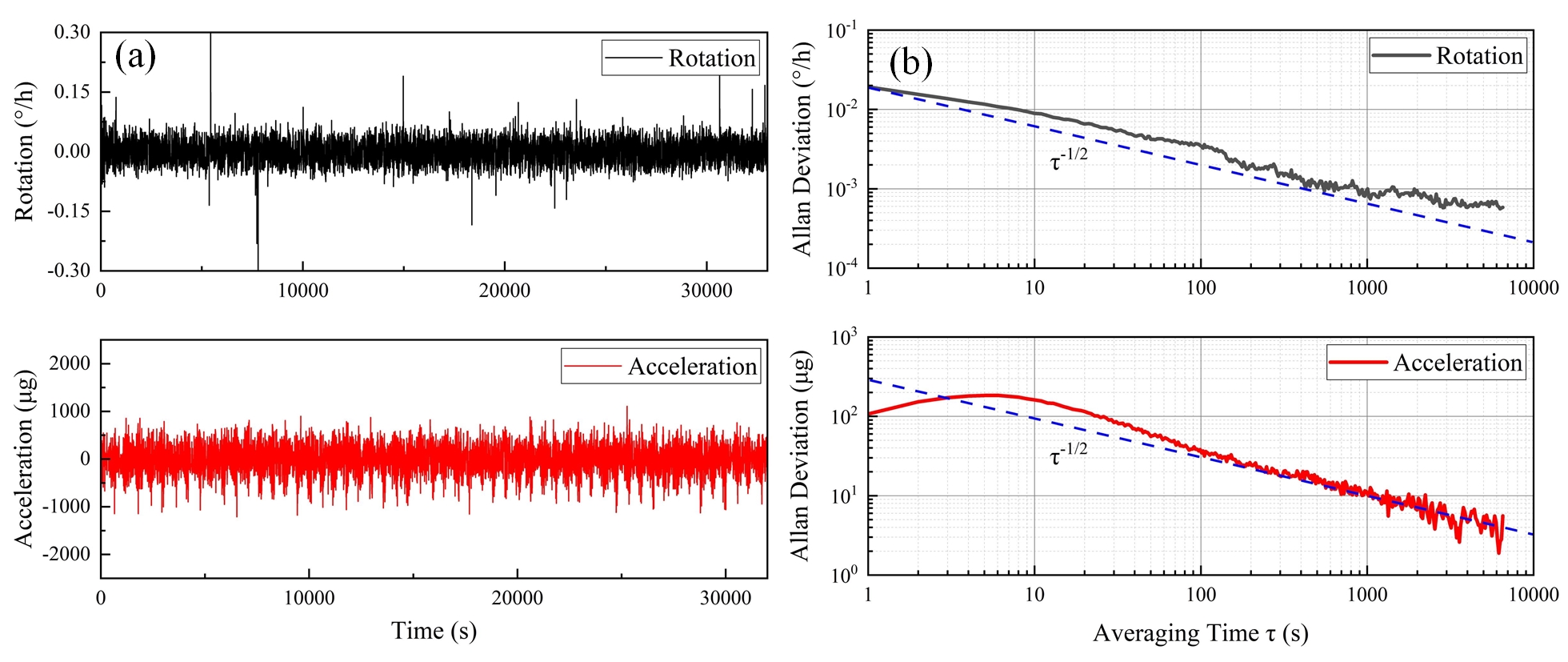}
    \caption{Long-term stability characterization of the dual-axis atomic interferometric inertial sensor. (a) Continuous rotation and acceleration data collected over 33,000 seconds. (b) Allan deviation of the measured signals.}
    \label{fig:Allan}
\end{figure*}

To enhance the long-term stability of the sensor, an area-reversal technique \cite{durfee2006} was adopted to cancel non-inertial phase shifts that are insensitive to the direction of the Raman effective wave vector. A real-time temperature compensation algorithm based on a Kalman Filter \cite{feng2015adaptive,cheiney2018navigation} was further employed to suppress the influence of temperature fluctuations, enabled by a distributed temperature monitoring system incorporating 20 sensors across the apparatus.

Figure \ref{fig:Allan} illustrates the long-term performance of the dual-axis sensor. In Fig. \ref{fig:Allan}(a), continuous rotation and acceleration data were collected over 33,000 seconds and used to calculate the Allan deviation shown in Fig. \ref{fig:Allan}(b). The sensor’s angle random walk (ARW) and velocity random walk (VRW) can be estimated from the initial points of the corresponding Allan deviation curves, with values of $3\times10^{-4}\,^\circ/\rm{\sqrt{h}}$ and $107\,\rm{\mu g/\sqrt{Hz}}$, respectively. At an integration time of 1000 s, the sensor demonstrated a long-term stability of $9\times10^{-4}\,\rm{^\circ/h}$ for rotation and $10\,\rm{\mu g}$ for acceleration. The short-time rise in the Allan deviation for acceleration is attributed to environmental vibrations, which could be further suppressed using active vibration isolation techniques. At longer integration times, the Allan deviation exhibits a characteristic $\sqrt{\tau}$ dependence, indicating that performance is dominated by white noise.

\section{CONCLUSION}

We demonstrate a dual-axis atomic interferometric inertial sensor capable of simultaneously measuring rotation and acceleration. Utilizing two transversely cooled, counter-propagating atomic beams, the sensor ensures a high data rate with zero dead time. The Raman beams with a $27\,\rm{cm}$ separation in our M-Z-type interferometric configuration are aligned using elongated mirrors, achieving a fringe contrast of about $5.9\,\%$ and $3.7\,\%$. Applying a quadrature demodulation method to the interference signal, both phase and amplitude are simultaneously extracted from the output of the lock-in amplifier. Subsequently, the rotational and accelerational components are discerned by calculating the average and half-difference between the output phases of two interferometers. 

Under forced oscillation in different directions, the sensor's outputs are compared with those of classical sensors, confirming its capability to measure rotation and acceleration synchronously in dynamic environments. Furthermore, we calibrate the gyroscope’s scaling factor and characterized nonlinearity on a turntable. For more precise acceleration calibration, a commercial vibration platform or tilt table could be employed; however, such implementations are constrained by the current system's size and will only be feasible after further miniaturization and refinement.

The atomic inertial sensor demonstrates an ARW of $3\times10^{-4}\,^\circ/\rm{\sqrt{h}}$ for rotation and a VRW of $107\,\mu g\mathrm{/\sqrt{Hz}}$ for acceleration, with shot-noise-limited sensitivities of $4\times10^{-5}\,^\circ/\rm{\sqrt{h}}$ and $0.37\,\mathrm{\mu}g\mathrm{/\sqrt{Hz}}$ based on current fringe contrast and atomic flux. There is still room for improvement before reaching these shot-noise-limited sensitivities due to technical noise in the present system, which may originate from environmental vibrations, laser power and frequency noise, and other factors. Further improvement can be achieved by optimizing key parameters. Using designed narrower Raman beams (less than $200\,\mu\mathrm{m}$) will lead to a enhancement in performance.Optimizing the atomic beam properties, such as narrowing the atomic beam linewidth by reducing the transverse temperature or employing transverse trapping, and increasing the atomic flux, the signal-to-noise ratio of the interference signal can be significantly improved. Additionally, fine-tuning the alignment and parallelism of the two atomic beams helps achieve common-mode noise suppression, thereby enhancing the performance of inertial measurement.

The dual-axis inertial sensor reaches a long-term stability of $9\times10^{-4}\,\rm{^\circ/h}$ for rotation and $10\,\rm{\mu g}$ at an integration time of 1000s. Further improvements in long-term performance can be achieved by enhancing active vibration isolation, optimizing the microwave and optical system stability, and implementing real-time bias estimation and correction algorithms. Under potential shot-noise-limited sensitivity, the corresponding long-term stability is anticipated to be $7.2\times10^{-5}\,^{\circ}/\mathrm{h}$ for rotation and $12\,\mathrm{ng}$ for acceleration
at the same integration time. These results demonstrate the potential of the sensor for long-duration, ultra-high-precision inertial navigation applications.

The ability to independently modulate each Raman beam in this configuration opens avenues for closed-loop inertial measurements in the upcoming phase of our research. The high data rate and bandwidth provided by the continuous atomic beam support the implementation of closed-loop control, which is expected to enhance scale-factor stability and expand the dynamic measurement range, promising greater adaptability to challenging environments compared to conventional cold atom interferometer inertial sensors \cite{sato2024closed}. 
While these attributes indicate potential for future development, translating the current performance into a practical IMU suitable for field applications still requires overcoming several substantial challenges. Key areas for improvement include conducting a systematic analysis of both short-term and long-term sensor behavior in static and dynamic conditions, investigating the thorough aspects of rotation and accelerometer biases, and compensate or mitigate uncertainties of alignment. Continued refinement in these areas, along with further experimental validation, is essential for achieving the required dynamic performance and long-term stability. These efforts will be critical to positioning the system as a viable building block for a complete quantum IMU.



%


\section*{Acknowledgments}
This work was supported by the National Natural Science Foundation of China (Grant No. 61473166).

\ifCLASSOPTIONcaptionsoff
  \newpage
\fi


\bibliographystyle{IEEEtran}






\end{document}